\documentclass[aps,prl,twocolumn,showpacs,amsmath,amssymb]{revtex4}
\usepackage{epsfig}
\usepackage{bm}

\begin{document}

\title{Counting statistics and detector properties of quantum 
point contacts}
\author{Dmitri V. Averin$^1$, and Eugene V. Sukhorukov$^2$}
\affiliation{$^1$\ Department of Physics and Astronomy, Stony Brook
University, SUNY, Stony Brook, NY 11794-3800 \\
$^2$\ D\'epartement de Physique Th\'eorique, 
Universit\'e de Gen\`eve, CH-1211 Gen\`eve 4,
Switzerland} 

\date{\today}

\begin{abstract}
Quantum detector properties of the quantum point contact (QPC)
are analyzed for arbitrary electron 
transparency and coupling strength to the measured system and 
are shown to be determined by the electron counting statistics. 
Conditions of the quantum-limited operation of the QPC detector 
which prevent information loss through the scattering time
and scattering phases are found for arbitrary 
coupling. We show that the phase information 
can be restored and used for the quantum-limited detection by 
inclusion of the QPC detector in the electronic Mach-Zehnder 
interferometer.
\end{abstract}

\maketitle

The dynamical process of quantum measurement is a crucial element 
of quantum-mechanical evolution \cite{b2,b3}. This fact should 
make it surprising that considerations related to quantum 
measurements do not typically play a role in the theory of 
mesoscopic conductors despite the quantum nature of electron 
transport in them. The main reason for this is that the dynamics 
of measurement does not affect single-time averages of 
individual operators that are typically studied in mesoscopic 
transport, and manifests itself explicitly only in the 
higher-order correlators, when the state obtained as a result 
of a measurement evolves and undergoes subsequent measurements. 
One example of this in mesoscopic transport is electron 
``counting statistics'' which determines full statistical 
properties of electron transfer through a conductor and 
requires an explicit discussion of the measurement procedure 
\cite{b4}. From the point of view of quantum measurement, 
the common assumption underlying different ways 
\cite{b3,b4} of obtaining this statistics is the 
statement that electron states in different electrodes 
of the structure after scattering represent classical, 
non-interfering events. The purpose of this work is to 
use this understanding to analyze one of the most basic 
mesoscopic detectors, the quantum point contact (QPC) beyond 
the linear-response regime and show that its properties are 
determined by the counting statistics. 

The QPC as the charge detector was suggested first in \cite{q1}, 
and studied experimentally in 
\cite{q2}. Currently, the QPCs are the main detectors used for 
measurements of the quantum-dot qubits \cite{q3,q31,q32,q34}. 
Theoretical description of the QPC detector has been worked out 
so far only in situations when the counting statistics does 
not play an essential role: in the ``tunnel-junction'' limit of 
small energy-independent transparency $D\ll 1$ \cite{q4,b8}, or for 
weak coupling to the measured systems \cite{q5,q6,q7,q8,q9,q10,q11}. 
Here we calculate the detector properties of the QPC for  
arbitrary energy-dependent transparency and for 
arbitrary coupling. Experimentally, this case is of interest since 
the typical measurements with QPC detectors are done in the regime 
$D \simeq 1/2$ which maximizes the detector sensitivity and for 
coupling that is not very weak \cite{q3}.

\begin{figure}
\epsfxsize=7.5cm
\epsfbox{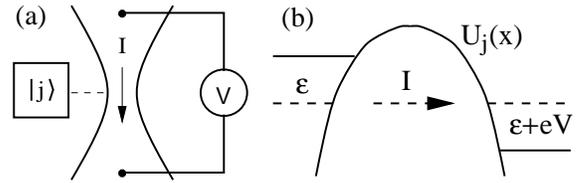}
\caption{(a) Real- and (b) energy-space diagram of the 
quantum-point-contact (QPC) detector. The scattering 
potential $U_j(x)$ for electrons in the QPC, and the 
current $I$ driven by the voltage $V$ are 
controlled by the state $|j\rangle$ of the measured system.}
\vspace{-5mm}
\end{figure}

The operating principle of the QPC detector is that the state 
of the measured system is reflected in the electron current 
$I$ between the QPC reservoirs (Fig.~1a). For the QPC to act 
as a detector, the spectral components of this current at 
relevant frequencies should be effectively classical, i.e., 
both the ``attempt frequency'' $eV/\hbar$, where $V$ the QPC 
bias voltage, and the inverse scattering time $\tau_0^{-1}$
should be much larger than the typical frequencies of the 
measured system. In this case, we can neglect the system 
dynamics and assume that its 
Hamiltonian is vanishing. It is convenient then to discuss the 
measurement in the basis of the eigenstates $|j\rangle$ of the 
system operator which couples it to the QPC, with each state 
producing its own scattering matrix $S_j$ for the QPC electrons.

{\em Measurement-induced dephasing}.---The basic physics of 
the measurement process is a trade-off between acquisition 
of information about the state of the measured system and 
back-action dephasing of this system (see, e.g., 
\cite{b8,q8,q9,q10}). For the quantum-limited detectors, 
the rates of the two processes coincide, for less efficient 
detectors the dephasing is more rapid than information gain. We 
begin by calculating the back-action dephasing rate for the QPC 
detector. Evolution in the scattering process of the total 
wavefunction $|\psi_0 \rangle$ comprised of the wavefunction 
$|in\rangle$ of an individual electron incident from one, say 
left, electrode, and the wavefunction of the measured system, 
$\sum_j c_j |j\rangle$, can be represented as  
\begin{equation}
|in\rangle \otimes \sum_j c_j |j\rangle \rightarrow 
\sum_j c_j\,\big(r_j |L\rangle + t_j |R\rangle\big)\otimes 
|j\rangle \, ,
\label{e1} 
\end{equation}
where $|L\rangle$ and $|R\rangle$ are the outgoing electron 
states in the left and right electrodes, respectively. The 
transmission and reflection amplitudes, $t_j$ and $r_j$, are 
the elements of the scattering matrix $S_j$: 
\begin{equation} 
S_j = \left( \begin{array}{cc} \displaystyle 
r_j\, , & \bar{t}_j  \\ t_j\, , & \bar{r}_j \, \end{array} 
\right)  
\label{e1*} 
\end{equation}
that depends on the state $|j\rangle$ of the measured system.
The resulting evolution of the density matrix $\rho$ of the 
system, $\rho= \mbox {Tr}_{el} |\psi_0 \rangle\langle \psi_0|$, 
is obtained by tracing over the electron states in Eq.\ 
(\ref{e1}): 
\begin{equation} 
\rho_{jk} =c_jc_k^* \rightarrow c_jc_k^* (t_jt_k^*+r_j 
r_k^*) \, .
\label{e2} 
\end{equation} 
Equation (\ref{e2}) shows that the absolute value of the 
off-diagonal elements $\rho_{jk}$, $j\neq k$, of $\rho$ are 
suppressed: $|t_jt_k^*+r_j r_k^*| \leq 1$, as a result 
of averaging over the two possible outcomes of scattering. 

We note that the process (\ref{e2}) 
requires that the state with energy $\epsilon$ in the left 
electrode is occupied with probability $f_1=f_F(\epsilon)$ 
(where $f_F$ is the Fermi function), while the corresponding 
state in the right electrode is empty with the probability 
$1-f_2$, where $f_2=f_F(\epsilon+eV)$ (see Fig.~1b). If both 
states are empty, there is no effect on the system density 
matrix. If both are occupied, they renormalize the energies 
of the measured system and therefore shift the phase of 
$\rho_{jk}$. Multiplying all types of evolution for a number 
$td\epsilon/ 2 \pi \hbar$ of different scattering events at 
energy intervals $d\epsilon$ during the time interval 
$t$ we find that the density matrix evolves
as $|\rho_{jk}(t)|= |\rho_{jk}(0)| e^{-\Gamma_{jk}t}$, 
where 
\begin{eqnarray} 
\Gamma_{jk}= &-& \int \frac{d\epsilon}{2 \pi \hbar} \ln | 
f_1f_2 e^{i \alpha} + f_1 (1-f_2) (t_jt_k^*+r_j r_k^*)  
\nonumber \\ &+&f_2 (1-f_1) (\bar{t}_j\bar{t}_k^* +\bar{r}_j 
\bar{r}_k^*)+ (1-f_1)(1-f_2)|\quad   
\label{e3} \end{eqnarray} 
is the dephasing rate.

The above heuristic derivation is presented to illustrate the 
measurement by QPC as the process of repeated interaction 
of the system with individual electrons and collection of 
statistics over the large number of scattering events, the 
concept that will be utilized below. Formally, the equation 
(\ref{e3}) can be obtained by starting with the expression for 
the time evolution of the density matrix $\rho$ due to 
interaction with the QPC: $\rho_{jk}(t)=\rho_{jk}(0) \langle 
e^{iH_kt} e^{-iH_jt} \rangle$, where $H_j$ is the Hamiltonian 
of the QPC electrons which includes the scattering potential 
$U_j(x)$ controlled by the measured system, and the average 
$\langle ... \rangle$ is taken over the stationary state of 
the QPC. Following the same steps as in the calculations of 
the counting statistics \cite{b4} one gets in the limit
$t>\hbar/eV$
\begin{equation}  
\frac{\rho_{jk}(t)}{\rho_{jk}(0)} = \exp\bigg\{ \int \frac{t 
d\epsilon}{2 \pi \hbar} \mbox{Tr} \ln [1-f +S_k^{\dagger} 
S_j f]\bigg\} \, , 
\label{e15} \end{equation} 
where $f$ is the matrix of occupation factors of the incident 
electron states: $f=\mbox{diag} \{ f_1,f_2 \}$. Evaluation of 
the trace in Eq.~(\ref{e15}) leads to Eq.~(\ref{e3}) with 
$e^{i \alpha} = \mbox{det} (S_k^{\dagger} S_j)$.

For a two-state system, there is only one rate  
$\Gamma_{12}\equiv \Gamma$, and Eq.\ (\ref{e3}) generalizes 
to arbitrary transmission properties of the QPC previous 
results for the ``back-action dephasing'' rate of the QPC 
detector obtained for energy-independent scattering 
amplitudes $t_j$, $r_j$. Indeed, in this case, and for 
vanishing temperature $T=0$, Eq.\ (\ref{e3}) gives  
\begin{equation}  
\Gamma = - (eV/2 \pi \hbar) \ln | t_1t_2^*+r_1 r_2^*|\, , 
\label{e4} 
\end{equation}  
the expression that reduces to the dephasing rate \cite{q4} 
$\Gamma = (eV/4\pi \hbar) (\sqrt{D_1}-\sqrt{D_2})^2$ 
in the tunnel limit, $D_j\equiv | t_j|^2 \ll 1$, or to the 
rate $\Gamma = (eV/2 \pi \hbar) (\Delta D)^2/[8D(1-D)]$
in the case of weak coupling 
$\Delta D  \equiv D_1-D_2 \ll (D_1+D_2)/2 \equiv D$ 
\cite{q5,q6}, 
if the coupling modulates only the absolute 
value of the transmission amplitude. 

{\em Detector counting statistics}.---The back-action dephasing
rates (\ref{e3}) characterize only one part of the dynamics 
of measurement, the other side being 
the information acquisition by the QPC detector. The information 
about the state $|j\rangle$ of the measured system is contained 
in the distribution of probabilities $P_{j}(n)$ for $n$ electrons 
to be transmitted through the QPC during the time interval $t$ 
if the system is in the state $|j\rangle$. With increasing $t$, the 
distributions $P_{j}(n)$ (and, accordingly, the states 
$|j\rangle$) can be distinguished with more and more certainty 
since the differences in their average positions and the widths 
grow, respectively, as $t$ and $t^{1/2}$. 

Quantitatively, the rates of information acquisition  
depend on the information measure used to 
characterize distinguishability of the distributions $P_j(n)$. 
In the context of 
quantum measurement, it is more appropriate \cite{b10,rem} to use 
for this purpose the measure given by the statistical overlap 
$M_{jk}(t)=\sum_n [P_j(n) P_k(n)]^{1/2} $, which is related 
to one of the R\'{e}nyi entropies \cite{b11} and not to the 
better-known Shannon entropy. At large time, $eVt/\hbar\gg 1$, 
when many electrons interact with the system, transport 
becomes a Markovian stochastic process. The generating functions 
of cumulants of $n$ are then proportional to time: 
$\ln\left[\sum_nP_{j}(n)\exp(\lambda n)\right]=t{\cal H}_j 
(\lambda)$. As a result, the overlap $M_{jk}$ exponentially 
decays in time,
\begin{equation}
M_{jk}(t)\equiv \sum_n [P_{j}(n) P_{k}(n)]^{1/2}=\exp[-W_{jk}t],
\label{e5} 
\end{equation}    
and one can introduce the notion of the measurement rate 
$W_{jk}$. A simple evaluation that involves the stationary 
point approximation expresses then the measurement rate in 
terms of the generating functions ${\cal H}_j(\lambda)$:
\begin{equation}
W_{jk}=-\,(1/2)\,{\rm min}_\lambda[{\cal H}_j(\lambda)+ 
{\cal H}_k(-\lambda)],
\label{rates} \end{equation}
which is so far a general result.

For a QPC, the counting statistics results \cite{b4}
can be summarized as follows. With probabilities $p_1=f_1 
(1-f_2)D_j(\epsilon)$ and $p_{-1}= f_2 (1-f_1) D_j(\epsilon)$ 
electrons are transmitted, respectively, forward and backward 
through the QPC changing $n$ by $\pm1$. With the probability
$p_0=1-p_1- p_{-1}$ the charge $n$ is not changed (when the 
states in both electrodes are simultaneously empty or occupied, 
or the incident electrons are reflected back into the 
electrodes). Using the expansion formula for the probabilities 
of a polynomial process, $(p_0+p_1+p_{-1})^N=\sum_{m_\pm} 
P(m_+,m_-)=1$, and inverting it to evaluate the moment 
generator  $\sum_{m_{\pm}}P(m_+,m_-)\exp[\lambda(m_+-m_-)]= 
[p_0+p_1e^{\lambda} +p_{-1}e^{-\lambda}]^N$ for the number of 
transmitted electrons $n=m_+-m_-$ per energy interval 
$d\epsilon=2\pi\hbar N/t$, we obtain the generating functions  
\begin{eqnarray}
{\cal H}_j({\lambda})=\int\frac{d\epsilon}{2 \pi \hbar}
\ln\big[1+f_1 (1-f_2) D_j(e^{\lambda}-1)\nonumber\\
+f_2 (1-f_1) D_j(e^{-\lambda}-1)\big],
\label{e6} 
\end{eqnarray} 
As shown below, Eq.~(\ref{e6}) means that 
$W_{jk}\leq \Gamma_{jk}$.

{\em Quantum limited detection}.---Equations (\ref{e3}), 
and  (\ref{e5})--(\ref{e6}) fully determine the detector 
properties of the QPC. In particular, they set the conditions 
for its ``ideal'' or quantum-limited operation characterized 
by the fact that the quantum coherence between the states 
$|j\rangle$ of the measured system is suppressed only by the 
acquisition of information about them, so that the measurement 
and dephasing rates coincide (see, e.g., Refs.\ \cite{b12}): 
\begin{equation}  
W_{jk}=\Gamma_{jk} \, . 
\label{e8} \end{equation} 
A first, obvious, requirement for this is sufficiently low 
temperature $T\ll eV$, when electrons pass through the QPC in 
only one direction, say from the left to right electrode. 
The next requirement follows from Eq.~(\ref{e3}) which 
shows that the dephasing rate $\Gamma$ can only increase 
without any changes in the probabilities $P_j(n)$, if the 
phases of the scattering amplitudes deviate from the relation  
\begin{equation} 
\phi_j=\phi_k\, , \qquad \phi_j \equiv \mbox{arg}(t_j/r_j) \, , 
\label{e9} \end{equation}   
under which they do not contain any information about the 
measured system. Similarly to the case of weak coupling 
\cite{q8,q9,q10}, an 
important example when Eq.~(\ref{e9}) is true is given 
by the symmetric potentials: $U_j(-x)=U_j(x)$. The unitarity 
of $S_j$ implies then that $\phi_j =\pi/2$. 

Finally, the last condition of the quantum limited operation 
restricts the energy dependence of the QPC transparencies 
$D_j(\epsilon)$. This condition is irrelevant when the bias 
is small, $eV \ll \tau_0^{-1}$, and $D_j$'s are effectively 
constant. In this case, the distributions $P_j(n)$ have 
the usual binomial form, $P(n)= C_N^nD^n R^{(N-n)}$, where 
$N=eVt/2 \pi \hbar$ and $R=1-D$. Taking the sum over $n$ 
explicitly in Eq.~(\ref{e5}) one sees then that if the other 
two ideality conditions are also satisfied, the QPC is the 
quantum-limited detector with 
\begin{equation} 
W_{jk} =\Gamma_{jk}=-\frac{eV}{2 \pi \hbar} \ln [ 
(D_jD_k)^{1/2} + (R_jR_k)^{1/2} ] \, .
\label{e7} \end{equation}  
In what follows we relax the two last requirements and show 
how the information contained in the scattering phases
can be extracted using tunable detectors, and how the energy 
dependence of the transmission can be optimized to reach the 
quantum-limited detection even at large voltages $V$. 

{\em Tunable quantum detector}.---If the condition (\ref{e9}) 
is not satisfied, the relative phases $\phi_j$ contain information 
about the quantum state that is lost in the measurement process. 
However, the fact that before the final state projection in the 
course of measurement the quantum coherence can be preserved 
and utilized leads 
to the following idea of a tunable detector: The information read 
via the interaction with the measured system, $|in\rangle \to 
S_j|\psi_j\rangle$, can be manipulated coherently, e.g. transported 
to a different location, before detection: $|\psi_j\rangle \to 
|\varphi_j\rangle =S_m|\psi_j\rangle$. These operations conserve 
the dephasing rate, $\langle\varphi_k|\varphi_j\rangle= \langle 
\psi_k|\psi_j \rangle$, but change the measurement rate, and 
can be used to tune the detector to the optimal point (\ref{e8}).  

\begin{figure}
\epsfxsize=5.5cm
\epsfbox{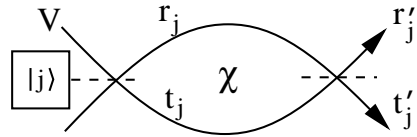}
\caption{The QPC detector included in the Mach-Zehnder 
interferometer: two chiral electronic modes mixing at the 
two QPCs (indicated by dashed lines) and accumulating the 
relative phase $\chi$ between them. The scattering amplitudes 
$t_j$, $r_j$ of one incident mode in the first QPC are 
controlled by the measured system. The interferometer 
enables one to use the phase information in these amplitudes
for detection.}
\vspace{-5mm}
\end{figure}

Here we consider the QPC detector included in the electronic 
Mach-Zehnder (MZ) interferometer \cite{b13} (Fig.~2) as an 
example of the set-up that provides such a new mode of operation. 
In contrast to the case of a single QPC, where the transmitted 
and reflected components of the electron wavefunction represent 
different classical outcomes, in the MZ interferometer, they 
can be mixed coherently at the next QPC after accumulating 
additional phase difference $\chi$ \cite{b14}. The phase 
information is converted in the second QPC into the transmission 
probabilities, and by adjusting $\chi$ and transmission $D_m$ of 
the second QPC, one should be able to achieve the 
quantum-limited detection even when Eq.~(\ref{e9}) is not true. 

Quantitatively, the transformation of the amplitudes $t_j$, 
$r_j$ in the output amplitudes $t_j'$, $r_j'$ (Fig.~2) is  
unitary: 
\begin{equation} 
S_m= 
\left(\begin{array}{cc} i (R_m)^{1/2} & (D_m)^{1/2}\,e^{-i\chi} 
\\ (D_m)^{1/2} & i (R_m)^{1/2}\,e^{-i\chi}  \end{array} \right).  
\label{e20} \end{equation}
For illustration purposes we assume the simplest case of 
the measured system having only two states, and of sufficiently 
small bias, $eV\tau_0\ll 1$, when transmission $D_m$ is 
energy-independent. If the measured system changes only the 
phases $\phi_j$ (\ref{e9}) of the scattering amplitudes, i.e.\ 
$D_1=D_2=D$, the single-QPC measurement rate vanishes. In 
contrast, the MZ interferometer converts the phase into the 
transmission probabilities and the measurement rate can be 
tuned to the maximum value (\ref{e8}) even in this case:
$W= \Gamma = -(eV/4\pi \hbar) \ln\left[1-4DR \sin^2\left( 
\phi/2)\right)\right]$, where $\phi=\phi_1-\phi_2$. This  
happens at the optimal point  
\begin{equation} 
\chi=(\phi_1+\phi_2)/2 \, , \quad D_m=R_m = 1/2, 
\label{e21} \end{equation} 
where $\chi$ is fixed by Eq.~(\ref{e21}) only if $D\neq 
1/2$ and only up to the shift $\chi\to\chi+\pi$. For 
$D=1/2$, $\chi$ is arbitrary. 

In general, if $D_1\neq D_2$, it is also possible to find 
the operating point which turns interferometer in the 
optimal detector. For instance, in the linear response regime,  
when $\Delta D=(D_1-D_2)/2$ and $\phi$ are small, one can see 
from Eqs.~(\ref{e20}) and (\ref{e7}) that this happens when 
the phase $\chi$ is given by Eq.~(\ref{e21}) and 
\begin{equation} 
D_m=1/2-(\Delta D/2)[(\Delta D)^2+\phi^2 (DR)^3]^{-1/2}. 
\label{e22} 
\end{equation}
This equation describes the transition from a maximally-mixing 
second QPC for $\phi \gg \Delta D$, as in Eq.~(\ref{e21}), when 
the measurement information is contained mostly in the phase, 
to the situation with no mixing for $\phi \ll \Delta D$, 
when from the very beginning the information is already 
in the transmission probabilities. 

{\em Optimization in energy space}.---If the bias voltage is not 
small, $eV\tau_0 \simeq 1$, the condition on $D_j(\epsilon)$ for 
the quantum-limited operation is found as follows. Eqs.\ 
(\ref{rates}) and (\ref{e6}) imply that the measurement rate $W$ 
is obtained by minimizing the function $\int d\epsilon \ln[(R_j+ 
D_je^{\lambda})(R_k+ D_ke^{-\lambda})]$ with respect to the 
variable $\lambda$. The dephasing rate ({\ref{e3}) can be also 
written as a minimum of the same function but taken locally for 
every energy $\epsilon$. Therefore $W_{jk}\geq \Gamma_{jk}$ with 
equality reached when the ratio $D_j(\epsilon)/R_j(\epsilon)$ has 
the same energy dependence in all states $|j\rangle$, differing 
only by the overall scale factor $C_j$: 
\begin{equation}
D_j(\epsilon)/R_j(\epsilon)= C_j D(\epsilon)/R(\epsilon) \, .
\label{e11} 
\end{equation} 
This equation determines when the QPC is still the 
quantum-limited detector (\ref{e8}) even for large bias $V$. It 
generalizes to arbitrary coupling strength the results of the 
linear-response theory \cite{q9,q10} obtained when the measured 
system simply shifts the energy of the QPC electrons. Indeed, 
in this case Eq.~(\ref{e11}) reduces to $D'(\epsilon) \propto 
D(\epsilon)R(\epsilon)$ in agreement with Refs.~\cite{q9,q10}. 
  
In the tunnel limit $D_j \ll 1$, or for weak backscattering 
$R_j \ll 1$, Eq.~(\ref{e11}) simplifies, respectively, to 
\begin{equation} 
D_j(\epsilon)= C_j D(\epsilon) \quad \mbox{or} \quad 
R_j(\epsilon)= C_j^{-1} R(\epsilon) \, .
\label{e12} 
\end{equation}  
These conditions have a simple interpretation. The energy 
dependence of $D_j(\epsilon)$ or, respectively, $R_j(\epsilon)$ 
determines the shape and, most importantly, the time delay 
of the wavepackets that correspond to individual electrons 
(holes) transmitted through the QPC. Equation (\ref{e12}) 
ensures that no information about the system is contained in 
the shape or position of these wavepackets that would be lost 
in the measurement process which is sensitive only to the 
transmission probabilities.

In conclusion, we have analyzed the detector properties of 
the QPC beyond the linear-response regime and found that 
both the back-action dephasing rate $\Gamma$ and the 
measurement rate $W$ are determined by the electron counting 
statistics. While generally $\Gamma\geq W$, the quantum-limited 
detection, $\Gamma=W$, can be reached by using tunable detectors 
which fully utilize the information in the scattering phases, 
and by optimizing the energy dependence of the transmission 
probabilities. 

\vspace*{.4ex}

We thank M. Buttiker and A.\ Jordan for discussions. This work was 
supported in part by the NSF under grant \# DMR-0325551 and by 
ARDA and DOD under the DURINT grant \# F49620-01-1-0439 (D.V.A.) 
and by the Swiss NSF (E.V.S.).


\begin{references}

\bibitem{b2} V.B. Braginsky and F.Ya. Khalili, {\em Quantum 
measurement\,}, (Cambridge, 1992). 

\bibitem{b3} {\em ``Quantum noise in mesoscopic physics''}, 
Ed.\ by Yu.V. Nazarov, (Kluwer, 2003). 

\bibitem{b4} L.S. Levitov, H. Lee, and G.B. Lesovik, J.\ Math.\ 
Phys. {\bf 37}, 4845 (1996). 

\bibitem{q1} M. Field {\em et al.},
%C.G. Smith, M. Pepper, D.A. Ritchie,
%J.E.F. Frost, G.A.C. Jones, and D.G. Hasko, 
Phys.\ Rev.\ Lett.
{\bf 70}, 1311 (1993).

\bibitem{q2} E. Buks {\em et al.}, 
%R. Schuster, M. Heiblum, D. Mahalu, V. Umansky, 
Nature {\bf 391}, 871 (1998).

\bibitem{q3} J.M. Elzerman {\em et al.}, 
%R. Hanson, J.S. Greidanus, L.H. Willems van Beveren, 
%S. De Franceschi, L.M.K. Vandersypen, S. Tarucha, and 
%L. P. Kouwenhoven, 
Phys.\ Rev. B {\bf 67}, 161308(R) (2003).

\bibitem{q31} T. Hayashi {\em et al.}, 
%T. Fujisawa, H.D. Cheong, Y.H. Jeong, and Y. Hirayama, 
Phys.\ Rev.\ Lett. {\bf 91}, 226804 (2003). 

\bibitem{q32} J.R. Petta {\em et al.}, 
% A.C. Johnson, C.M. Marcus, M.P. Hanson, and A.C. Gossard, 
Phys.\ Rev.\ Lett. {\bf 93}, 186802 (2004). 

\bibitem{q34} A.K. Huettel {\em et al.}, 
%S. Ludwig, K. Eberl, J.P. Kotthaus, 
cond-mat/0501012. 

\bibitem{q4} S.A. Gurvitz, Phys.\ Rev. B {\bf 56}, 15215 (1997).

\bibitem{b8} A.N. Korotkov, in Ref.\ \onlinecite{b3}, p.\ 205. 

\bibitem{q5} I.L. Aleiner, N.S. Wingreen, and Y. Meir, Phys.\ 
Rev.\ Lett. {\bf 79}, 3740 (1997).

\bibitem{q6} Y. Levinson, Europhys.\ Lett. {\bf 39}, 299 (1997).

\bibitem{q7} L. Stodolsky, Phys.\ Rep. {\bf 320}, 51 (1999).

\bibitem{q8} D.V. Averin, in: {\em ``Exploring the 
quantum/classical frontier''}, Ed.\ by J.R. Friedman and S. Han, 
(Nova Science Publishes, NY, 2003), p.\ 447; cond-mat/0004364.

\bibitem{q9} S. Pilgram and M. B\"{u}ttiker, Phys.\ Rev.\ Lett.
{\bf 89}, 200401 (2002).

\bibitem{q10} A.A. Clerk, S.M. Girvin, and A.D. Stone, Phys.\ 
Rev. B {\bf 67}, 165324 (2003). 

\bibitem{q11} A.N. Jordan and M. Buttiker, Phys.\ Rev. B {\bf 
71}, 125333 (2005); cond-mat/0505044.   

\bibitem{b10} W.K. Wootters, Phys.\ Rev. D {\bf 23}, 357 (1981).

\bibitem{rem} This conclusion may also follow from the consideration
of the conditional evolution \cite{b8} of the measured system.  

\bibitem{b11} C. Beck and F. Schl\"{o}gl, {\em Thermodynamics 
of chaotic systems\,}, (Cambridge, 1993). 

\bibitem{b12} D.V. Averin, in Ref.\ \onlinecite{b3}, p.\ 229. 

\bibitem{b13} Y. Ji {\em et al.}, 
%Y. Chung, D. Sprinzak, M. Heiblum, D. Mahalu, and H. Shtrikman, 
Nature {\bf 422}, 415 (2003). 

\bibitem{b14} P. Samuelsson, E.V. Sukhorukov, and M. Buttiker, 
Phys.\ Rev.\ Lett. {\bf 92}, 026805 (2004).

\end{references}
\end{document}